\documentclass[aps,showpacs,amsmath,amssymb,twocolumn,pra,superscriptaddress,notitlepage]{revtex4-2}

\usepackage{qcircuit}
\usepackage{amsmath,bm}
\usepackage[dvips]{graphicx}
\usepackage{amsmath,amssymb,amsthm,mathrsfs,amsfonts,dsfont}
\usepackage{subfigure, epsfig}
\usepackage{braket}
\usepackage{bm}
\usepackage{enumerate}
\usepackage{color}
\usepackage{graphicx}
\usepackage{algorithm}
\usepackage{algorithmic}
\usepackage{braket}
\usepackage{comment}
\usepackage{appendix}
\usepackage{here}
\usepackage{tabularx}
\usepackage{physics}
\usepackage{mathtools}

\begin{document}


\title{Projective squeezing for translation symmetric bosonic codes}


\author{Suguru Endo}
\email{suguru.endou@ntt.com}
\affiliation{NTT Computer and Data Science Laboratories, NTT Corporation, Musashino, 180-8585, Tokyo, Japan}

\author{Keitaro Anai}
\affiliation{%
 Department of Applied Physics, School of Engineering, The University of Tokyo,\\
 7-3-1 Hongo, Bunkyo-ku, Tokyo 113-8656, Japan
}%

\author{Yuichiro Matsuzaki}
\affiliation{Department of Electrical, Electronic, and Communication Engineering, Faculty of Science and Engineering, Chuo University, 1-13-27, Kasuga, Bunkyo-ku, Tokyo 112-8551, Japan}

\author{Yuuki Tokunaga}
\affiliation{NTT Computer and Data Science Laboratories, NTT Corporation, Musashino, 180-8585, Tokyo, Japan}

\author{Yasunari Suzuki}
\affiliation{NTT Computer and Data Science Laboratories, NTT Corporation, Musashino, 180-8585, Tokyo, Japan}


\begin{abstract}

 The design of translation symmetric bosonic codes, e.g., Gottesmann-Kitaev-Preskill and squeezed cat codes, is robust against photon loss, but the computation accuracy is limited by the available squeezing level. Here, we introduce the \textit{projective squeezing} (PS) method for computing outcomes for a higher squeezing level by revealing that a linear combination of displacement operators with periodic displacement values constitutes the smeared projector onto the better code space. We also show the analytical relationship between the increased squeezing level and the projection probability. We introduce concrete implementation methods for PS based on linear-combination-of-unitaries (LCU) and virtual quantum error detection (VQED) methods. \textcolor{black}{While both methods involve an ancilla qubit, the VQED method can be performed by much shallower depth and offer a noise-robust implementation against the ancilla qubit noise.}

\end{abstract}

\maketitle
\emph{Introduction.---} 
\label{section: introduction}
Quantum error correction (QEC) is a necessary building block to protect quantum information from the effects of noise for realizing quantum computing~\cite{devitt2013quantum,lidar2013quantum,nielsen2010quantum}. Bosonic quantum codes are promising candidates for QEC because they are constructed by extracting the valuable subspace for encoding from the infinite dimensions of a harmonic oscillator and, hence, are hardware efficient~\cite{cai2021bosonic,terhal2020towards,joshi2021quantum}. Bosonic quantum codes can be characterized by their symmetries, e.g., cat and binomial codes have rotation symmetries~\cite{cochrane1999macroscopically,michael2016new,grimsmo2020quantum}. Due to this symmetry, higher-order rotation codes than four-legged rotation codes have error correction properties for photon loss~\cite{ofek2016extending,leghtas2013hardware}. Another remarkable property of rotation codes is that two-legged cat codes have a biased property for bit and phase flip errors; as the amplitude of the cat states increases, phase errors are mitigated, with bit-flip errors being dominant~\cite{grimm2020stabilization,lescanne2020exponential}.

Meanwhile, Gottesman-Kitaev-Preskill (GKP) codes have translation symmetries on the phase space~\cite{gottesman2001encoding,grimsmo2021quantum}. Notably, GKP codes outperform rotation symmetric codes in quite broad parameter regimes for photon loss errors, although they have been designed to suppress displacement errors~\cite{albert2018performance}. This is because the displacement stabilizer operators of GKP can protect the state from displacement operators that expand the errors. Recently, squeezed cat (SC) codes, constructed from a superposition of squeezed coherent states with opposite amplitudes, have been studied as an alternative approach for QEC~\cite{schlegel2022quantum,xu2023autonomous,hillmann2023quantum}. SC codes have translation symmetries as well as rotation symmetries. In the SC code, dephasing errors are suppressed as the squeezing parameter increases due to the orthogonality of code states, while the effect of bit-flip errors due to photon loss is alleviated by keeping the amplitudes small. Although translation symmetric codes such as the GKP and the SC codes have their advantages, the orthogonality of the logical states is restricted by the experimentally realizable squeezing level~\cite{eickbusch2022fast,campagne2020quantum,sivak2023real,pan2023protecting}, which leads to a finite computation error due to measurement and gate errors~\cite{gottesman2001encoding,royer2020stabilization,hastrup2021improved}. See Supplementary Materials (SM) \footnote{Supplementary Materials} for logical Pauli measurement errors in SC codes.

\begin{figure*}[t!]
    \centering
    \includegraphics[width=2\columnwidth]{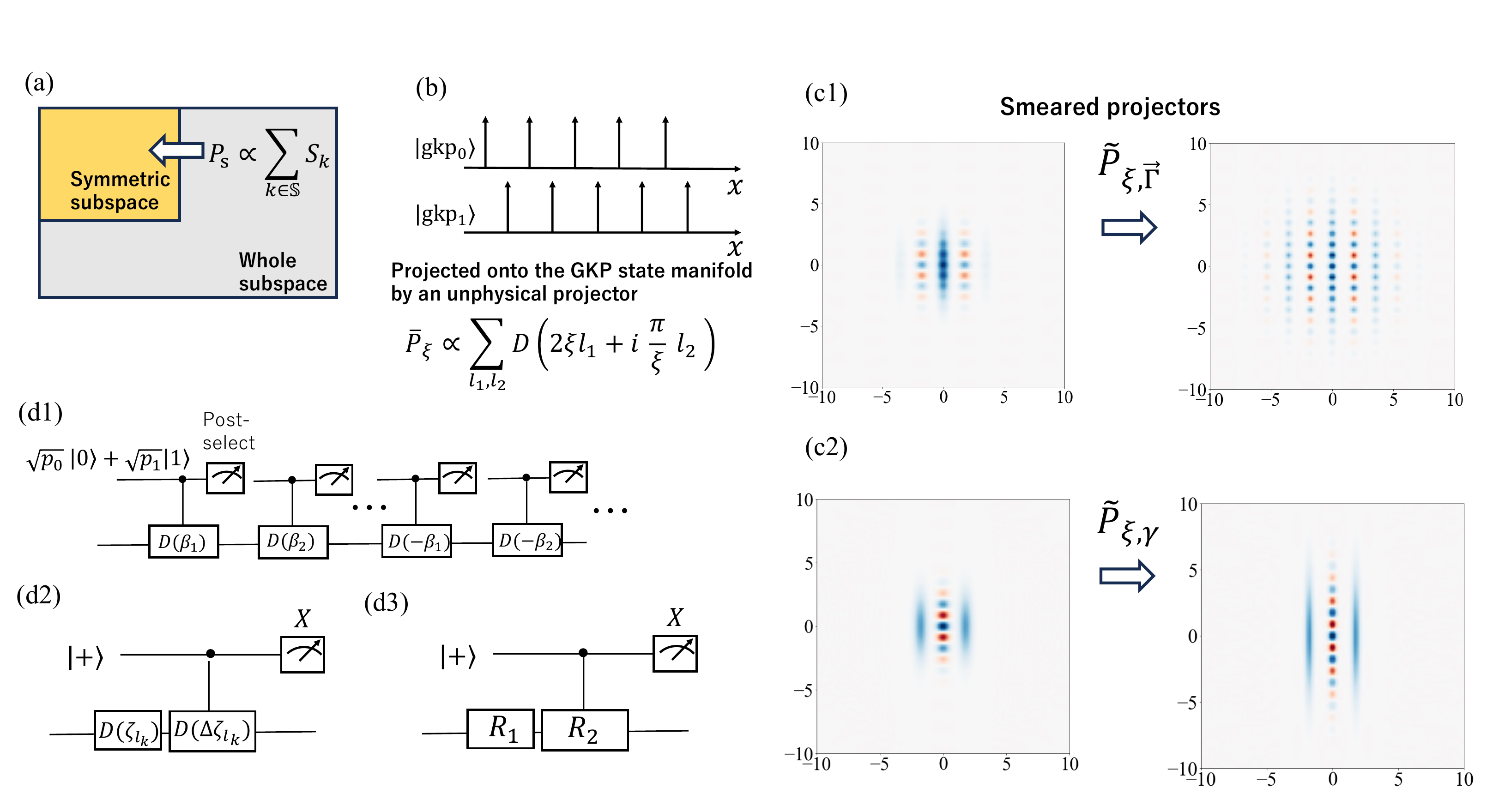}
    \caption{The schematic of our protocol to project the quantum states onto the translation symmetric subspace. (a) Projection onto the symmetric subspace. In the presence of symmetry in the system, we can construct the projector onto the symmetric subspace as a linear combination of symmetry operators, e.g., stabilizer operators of quantum error correction codes. (b) An example of quantum states with translation symmetry. The ideal GKP states have translational symmetry, and their projectors can be constructed from the uniform summation of an infinite number of stabilizer displacement operators. Note that this projection cannot be performed in reality due to the unphysicality of the ideal GKP states. (c) Squeezing by the projection onto the translation symmetric subspace. We construct the approximate projector as a linear combination of displacement operators with Gaussian weight in accordance with the displacement amount, as described in the main text. This work mainly considers GKP (c1) and SC states (c2). (d) Quantum circuits introduced in this work. (d1) The LCU circuit to project the state onto a higher squeezing level subspace, which requires repetitive applications of controlled-displacement operations and post-selection. (d2) Quantum circuit for \textit{virtually} squeezing quantum states, which requires a much shorter quantum circuit with a constant depth. For $\zeta_{l_k}$ and $\Delta \zeta_{l_k}$, see the implementation part of the virtual PS in the main text. (d3) We can also add rotation degrees of freedom to further suppress errors. }
    \label{Fig:vqsd}
\end{figure*}

In the present paper, we propose the \textit{projective squeezing} (PS) method for obtaining the computation results corresponding to higher squeezing levels at the cost of sampling overheads in accordance with the increased squeezing level. The schematic of our protocol is shown in Fig. \ref{Fig:vqsd}. By linearly combining the stabilizer operators, we can constitute the projector onto the symmetric subspace (Fig. \ref{Fig:vqsd} (a)) \cite{cai2021quantum}. However, the projector onto the ideal translation symmetric manifold is unphysical (Fig. \ref{Fig:vqsd} (b)).   
Thus, we first show that the linear combination of displacement operators with Gaussian weight according to the displacement value can constitute the physical \textit{smeared} projector that projects the state onto a higher squeezing level subspace (Fig. \ref{Fig:vqsd} (c)). We give a clear analytical expression of the tradeoffs between the increased squeezing level and the projection probabilities.

Then, we show that the PS protocol can be performed in two ways. The first one uses the linear-combination-of-unitaries (LCU) method~\cite{childs2012hamiltonian,low2019hamiltonian}. Although LCU generally employs  $\mathrm{log}(N_{\rm LCU})$ ancilla qubits and controlled operations from them for the number of unitaries $N_{\rm LCU}$, we show that we can apply the smeared projector with repetitive applications of LCU circuits involving only an ancilla qubit and a controlled-displacement operation. The second method is the virtual PS method. The word \textit{virtual} indicates that we can obtain expectation values of observables for more ideal states, but not quantum states themselves, by post-processing measurement outcomes in a similar spirit to quantum error mitigation (QEM)~\cite{li2017efficient,temme2017error,endo2018practical,cai2023quantum,endo2021hybrid}. We generalize the virtual quantum error detection (VQED) method~\cite{bonet2018low,PhysRevA.108.042426,mcclean2020decoding,cai2021quantum} for projecting the state onto the translation symmetric subspace. The required depth of the quantum circuit for implementation is much shallower than the LCU implementation. \textcolor{black}{In addition, while this method also involves an ancilla qubit, we show that we can modify the VQED strategy to construct the unbiased estimator of the noiseless computation result.} We also confirm our results via numerical simulations and show that our protocol can suppress the photon loss errors.

\emph{Translation symmetric bosonic codes.---}  Let the displacement operator and the squeezing operator denote as $D(\xi)=\mathrm{exp}(\xi a^\dag-\xi^* a) $and $S(z)=\mathrm{exp}(\frac{1}{2}(z^* a^2-z a^{\dag 2} )~(\xi, z \in \mathbb{C})$ ~\cite{gerry2023introductory,klauder2006fundamentals} . The SC code states are the superposition of the SC states represented as $
\ket{\rm{sq}_{\xi, z}^{\mu}}= \frac{1}{N_{\rm sq}}(\ket{\xi,z} +(-1)^\mu \ket{-\xi,z})$~\cite{schlegel2022quantum,xu2023autonomous,hillmann2023quantum}, where the squeezed coherent state reads $\ket{\xi,z} \equiv D(\xi) S(z) \ket{0}~(\xi\in \mathbb{R}, z>0)$~\cite{gerry2023introductory} and $N_{\rm sq}$ is the normalization factor. They have a translation symmetry toward the momentum axis, as the periodic interference fringe structure in the Wigner function in Fig. \ref{Fig:vqsd} (c2) shows. As we detail in SM, the logical X operation for the SC code can be performed with $X_{\rm{sq}} \equiv -i D(i\frac{\pi}{4\xi} )$, and the code states are stabilized by $X_{\rm{sq}}^2=-D(i\frac{\pi}{2\xi} )$. This indicates that the SC code states have an approximate translation symmetry with the displacement operators $\{ (-1)^m D(i (\frac{\pi}{2\xi}) m)  \}_m ~(m \in \mathbb{Z})$. SC codes have been pointed out to be potentially robust to both dephasing and photon loss errors by investigating Knill-Lalamme error correction conditions~\cite{schlegel2022quantum}. 

Another important class of the translation symmetric bosonic codes is the GKP codes, which have a translational symmetry for two distinct directions on the phase space~\cite{grimsmo2021quantum,gottesman2001encoding}. The ideal, and at the same time unphysical GKP logical zero states with the rectangular lattice structure are described with the superposition of the infinitely squeezed coherent states as $\ket{\mathrm{gkp}_0}=\sum_n \ket{2 n \xi, z } $ with the logical one state being $\ket{\mathrm{gkp}_1}=\sum_n \ket{(2n+1) \xi, z } $ for $z \rightarrow \infty$. These logical states have the translation symmetry both in the momentum and the position directions, i.e., they are stabilized by the displacement operators $S_Z=D(i \frac{\pi}{\xi})$ and $S_X=D(2 \xi)$. For the square-structure GKP code, we have $\xi= \sqrt{\pi/2}$. Since the ideal GKP states are unphysical, we need to introduce the approximate GKP states defined as $\ket{\rm{gkp}^{\Delta}_\mu}=\frac{1}{N_{\rm gkp}^{\Delta}} e^{-\Delta^2 a^\dag a}  \ket{\rm{gkp}_\mu}$, where $N_{\rm gkp}^{\Delta}$ is the normalization factor. Note that the factor of $\Delta^2$ is relevant to the squeezing level, and $\Delta \rightarrow 0$ corresponds to the superposition of the infinitely squeezed coherent states. The GKP states are shown to be robust to photon loss errors theoretically in broad regimes, compared with rotation symmetric bosonic codes~\cite{albert2018performance}, and experimentally achieve the break-even point~\cite{sivak2023real}, where the coherence time of error-corrected qubits exceeds that of physical qubits. 

\emph{Projective squeezing.---}
The experimentally available squeezing level restricts the computation accuracy. Here, to alleviate such problems, we propose the \textit{projective squeezing} (PS) technique, which allows us access to computation results for higher squeezed levels at the expense of the sampling overhead. We first consider the PS for the squeezed coherent cat states. While we can construct the projector onto the code subspace as $P_{\rm code}=\frac{1}{|\mathbb{S}|}\sum_k S_k$ for the set of stabilizers $\mathbb{S}$ and its elements $S_k \in \mathbb{S}$~\cite{cai2021quantum,mcclean2020decoding}, the number of stabilizers for translation symmetric codes is infinite and projection probability onto ideally translation symmetric subspace is infinitely small. Then, we introduce the following smeared projector that projects the finite-level SC states onto the higher squeezing level subspace: $\tilde{P}_{\xi, \Gamma} = \sum_l p_l^{\xi, \Gamma} (-1)^l D\big(i \frac{\pi}{2 \xi} l\big), p_l^{\xi, \Gamma} \propto e^{-\frac{1}{\Gamma^2} (\frac{\pi}{2 \xi} l)^2}, \sum_l p_l^{\xi, \Gamma}=1.$ Here, we set $\Gamma^2 =e^{2 z} (e^{2\delta z} -1)$. Then, we analytically show that 
\begin{equation}
\begin{aligned}
\frac{1}{\sqrt{q_{z, \delta z}}}\tilde{P}_{\xi, \Gamma} \ket{\mathrm{sq}_{\xi, z}^{\mu}} &\sim \ket{\mathrm{sq}_{\xi, z+\delta z}^{\mu}} \\
q_{z, \delta z} &\sim e^{-\delta z},
\label{Eq: prosq}
\end{aligned}
\end{equation}
where $q_{z, \delta z}=  \bra{\mathrm{sq}_{\xi, z+\delta z}^{\mu}}\tilde{P}_{\xi, \Gamma}^\dag \tilde{P}_{\xi, \Gamma} \ket{\mathrm{sq}_{\xi, z+\delta z}^{\mu}} $ is the projection probability. While we approximate the summation with the Gaussian integral, when the Gaussian is sufficiently narrower than the summation interval, it leads to the approximation error. Therefore, the approximation in Eq. \eqref{Eq: prosq} is quite accurate when $e^{2 z} \gtrsim (\frac{\pi}{2 \xi})^2$ and $\delta z \gtrsim \frac{1}{2} e^{-2 z} (\frac{\pi}{2 \xi} )^2$. See SM for the calculation details. Note that this result can be straightforwardly extended to squeezed comb states that are defined as the superposition of equally spaced squeezed coherent states in the position axis~\cite{shukla2021squeezed}.

We can also have a similar argument for the rectangular structure GKP states. Because the GKP states have translation symmetries in both position and momentum direction, we now define the smeared projector as $\tilde{P}_{\xi, \vec{\Gamma}}=\sum_{\vec{l}} p_{\vec{l}}^{\xi, \vec{\Gamma}} D\left(2\xi l_1 +i \frac{\pi}{\xi}l_2\right), 
p_{\vec{l}}^{\xi, \vec{\Gamma}} \propto e^{-\frac{1}{{\Gamma_1}^2} (2 \xi l_1)^2} e^{-\frac{1}{{\Gamma_2}^2} (\frac{\pi}{\xi} l_2)^2}, \sum_{\vec{l}} p_{\vec{l}}^{\xi, \vec{\Gamma}} =1$
with $\vec{l}=(l_1, l_2) \in \mathbb{Z}^2$ and $\vec{\Gamma}=(\Gamma_1, \Gamma_2) \in \mathbb{R}^2$. For $\Gamma_1=\Gamma_2=\Gamma_0$, the application of the smeared projector onto the square GKP states $\ket{\mathrm{gkp}_\mu^{\Delta}}$ yields
an approximation of a square GKP state for the damping parameter $\Delta'$ such that $\frac{1}{\Delta'^2}=\frac{1}{\Delta^2}+\Gamma_0^2$. By setting $\Gamma_0^2=\frac{1}{\Delta^2}(s-1) ~(s>1)$, we obtain
\begin{equation}
\begin{aligned}
\frac{1}{\sqrt{q_{\Delta, s}}}\tilde{P}_{\xi, \Gamma} \ket{\mathrm{gkp}_\mu^{\Delta}}  &\sim  \ket{\mathrm{gkp}_\mu^{ \Delta s^{-\frac{1}{2}}}  }  \\ 
q_{\Delta, s} & \sim s^{-1},
\end{aligned}
\end{equation}
for the projection probability is $q_{\Delta,s}=\bra{\mathrm{gkp}_\mu^{\Delta}} \tilde{P}_{\xi, \vec{\Gamma}}^\dag \tilde{P}_{\xi, \vec{\Gamma}}  \ket{\mathrm{gkp}_\mu^{\Delta}} $. The approximation error arises due to the similar procedure for the SC code; the approximation is valid for $s \gtrsim 1+ 2 \xi \Delta^2$.
Refer to SM for the detailed derivation. From the relation of $z= -\mathrm{ln} \Delta$~\cite{albert2018performance}, we have $s=e^{ 2 \delta z}$, which results in $q_{\Delta,s}=q_{\delta z}^2$. This is because PS for the GKP requires projection towards the position and momentum axes.

We numerically show that projection onto the higher squeezing subspace can mitigate the effect of photon loss errors, and our method is also compatible with the projection onto the rotation symmetric subspace, rendering the quantum computation more robust to the photon loss errors \cite{endo2022quantum} because rectangular-structure GKP code states and SC $0/1$ states also have rotation symmetries. See SM for more details.

\emph{LCU projective squeezing. ---} We here discuss the method for acquiring the physical state with a higher squeezing level by using an LCU algorithm~\cite{childs2012hamiltonian,low2019hamiltonian}. Although the LCU algorithm was shown to be useful for state preparation of the bosonic and spin-ensemble GKP states in Ref. \cite{omanakuttan2023spin}, we also reveal that it can be leveraged for squeezing for the GKP and the SC code states. We prepare the qubit-based ancilla state $\ket{\phi_{\rm anc}}=G_{\rm prep} \ket{{0}}^{\otimes N_{\rm anc}}= \sum_{l=1}^d \sqrt{p_l} \ket{l}$ with $G_{\rm prep}$ being the state-preparation oracle unitary by using a $N_{\rm anc}=\lceil \mathrm{log}_2 (d) \rceil$ qubits and the select unitary $U_{\rm sel}= \sum_{l=1}^n \ket{l}\bra{l} \otimes h_l D(\zeta_l)$. Here, $h_l$ is the sign accompanying the displacement operator. Denoting the register state as $\ket{\psi}_{reg}$, we obtain  $\bra{\phi_{\rm anc}}U_{\rm sel} \ket{\phi_{\rm anc}} \otimes \ket{\psi_{\rm reg}} \propto \tilde{P} \ket{\psi_{\rm reg}}$
for $\tilde{P}=\sum_{l=1}^{d} p_l h_l D(\zeta_l) $. This transformation can be performed by applying the select unitary $U_{\rm sel}$ to the state $\ket{\phi_{\rm anc}} \otimes \ket{\psi_{\rm reg}}$ followed by $G_{\rm prep}^\dag$ and the projection to the ancilla initial state $\ket{{0}}^{\otimes N_{\rm anc}}$. The successful projection probability is $q=\mathrm{Tr}[\tilde{P} \rho_{\rm reg} \tilde{P}^\dag ]$ for the register density operator $\rho_{reg}$. The complex controlled operation is generally hard to implement for the near-term quantum hardware. To circumvent this problem, we consider the product form of the smeared projector, i.e., $\tilde{P}_{\rm pro}=(Q_0 Q_0^\dag)^M $ for the SC state and $\tilde{P}_{\rm pro}= (Q_1 Q_1^\dag Q_2 Q_2^\dag)^M $ for the GKP state with $M$ being the positive integer. Here, $Q_0=p_0 I- p_1 D(i \frac{\pi}{2 \xi})$, $Q_1= p_0 I + p_1 D(2 \xi)$  and $Q_2= p_0 I + p_1 D(i \frac{\pi}{\xi})$. Note that the product form approaches the Gaussian-weight projector for sufficiently large $M$ because the binomial distribution converges to the Gaussian distribution due to the central limit theorem~\cite{feller1991introduction}, with the width of the Gaussian distribution being $\Gamma^2=(\frac{\pi}{\xi})^2 M p_0 (1-p_0)$ for the SC code and $\Gamma_0^2= 8 \pi M p_0 (1-p_0)$ for the square GKP code. 
\textcolor{black}{
Let $T_{\rm LCU}$ denote the required gate times except for the measurement process where $T_{\rm LCU}$ is proportional to $M$.
}
\textcolor{black}{Because $\Gamma^2, \Gamma_0^2  \propto e^{2z} (e^{2 \delta z }-1)$, we have $T_{\rm LCU} = O (e^{2z} (e^{2 \delta z }-1))$. } Note that the projection via the product-form smeared projector $\tilde{P}_{\rm pro}$ can be performed by applying the repetitive LCU quantum circuits only using one ancilla qubit as shown in Fig. \ref{Fig:vqsd} (d1). We post-select the measurement outcome $1$ for the SC code due to the negative sign, while we post-select $0$ for the GKP state.

\emph{Virtual projective squeezing.---} Now, we introduce the implementation of the virtual PS protocol. The virtual PS method has a significant hardware advantage over the LCU implementation: it allows for the computation of expectation values corresponding to a higher squeezing level, only necessitating an ancilla qubit and a randomly generated controlled displacement operation. Unlike the LCU implementation, the circuit depth of virtual PS is constant, irrespective of the probability weight and the number of operators constituting the projector. In this method, we can compute expectation values of observables for the state as 
$\rho_{\rm{vs}}= \frac{\rho_{\rm{vs}}'}{\Tr[\rho_{\rm{vs}}']} $ for $\rho_{\rm{vs}}'= \tilde{\mathcal{P}}_N \circ \mathcal{U}_N \circ... \tilde{\mathcal{P}}_1 \circ \mathcal{U}_1 \circ    (\rho_{\rm{vac}}^{\otimes N_{\rm q}})$. Here, we assume we apply PS after each state preparation and gate operation, and $\rho_{\rm{vac}}$ is the vacuum state, $N_{\rm q}$  is the number of bosonic qubits, $\mathcal{U}_i ~(i=1, ..., N)$ are the gate operation processes including the state-preparation operations, and $\tilde{\mathcal{P}}_k (\cdot) = \tilde{P}_k (\cdot) \tilde{P}_k^\dag~(k=1, 2, ..., N)$ is the PS operation. We assume that $\tilde{\mathcal{P}}_k$ operates only on one bosonic qubit. By substituting $\tilde{P}_k = \sum_{l_k} p_{l_k} h_{l_k} D_k (\zeta_{l_k})$ with $h_{l_k}$ taking $h_{l_k}=1$ for the GKP case but taking $h_{l_k} \pm 1$ for the SC case, we obtain the expectation value for the observable $O$ for $\rho_{\rm vs}$ as 
\begin{equation}
\langle O \rangle _{\rm vs}=\frac{\sum_{\vec{l}, \vec{l}'} p_{\vec{l},\vec{l}'} \Tr[O \tilde{\mathcal{P}}_{l_N, l'_N} \circ \mathcal{U}_L\circ...\circ \tilde{\mathcal{P}}_{l_1, l'_1} \circ \mathcal{U}_1 (\rho_{\rm vac}^{\otimes N_{\rm q}}) ]}{\sum_{\vec{l}, \vec{l}'} p_{\vec{l},\vec{l}'} \Tr[\tilde{\mathcal{P}}_{l_N, l'_N} \circ \mathcal{U}_L\circ...\circ \tilde{\mathcal{P}}_{l_1, l'_1} \circ \mathcal{U}_1 (\rho_{\rm vac}^{\otimes N_{\rm q}}) ]}.
\label{Eq:vqd}
\end{equation}
Here, we denote $\tilde{\mathcal{P}}_{l_k, l_k'}(\cdot)=h_{l_k}h_{l'_k} D(\zeta_{l_k}) (\cdot) D^\dag(\zeta_{l'_k})$ and $p_{\vec{l},\vec{l}'}=\prod_{k=1}^N p_{l_k} p_{l'_k} $ for $\vec{l}=(l_{N},l_{N-1},...,l_1)$ and $\vec{l}'=(l'_{N},l'_{N-1},...,l'_1)$. Then, we can compute Eq. (\ref{Eq:vqd}) by evaluating the numerator and the denominator on the hybrid system of two-level system and harmonic oscillators separately and post-processing the result as follows: 

\begin{enumerate}
    \item For $n = 1,\cdots, L$, iterate the following steps:
    \begin{enumerate}
        \item Sample $\vec{l}$ and $\vec{l}'$ with the probability $p_{\vec{l}, \vec{l}'}$.
        \item Execute the quantum circuit shown in Fig. \ref{Fig:vqsd} (d2) in accordance with the sampled $\vec{l}$ and $\vec{l}'$ after each gate for computation. Here, $\Delta \zeta_{k}= \zeta_{l_k}- \zeta_{l'_{k}}$. 
        \item Store the product of the measurement outcomes of ancilla qubits as $m_n$ and $m_n'=m_n o_n$ for the measurement outcome of the observable $o_n$.
    \end{enumerate}
    \item Compute $\langle m \rangle = \frac{1}{L}\sum_{n=1}^L m_n$ and $\langle m' \rangle = \frac{1}{L}\sum_{n=1}^L m'_n$. 
    \item Output $\frac{\langle m' \rangle}{\langle m \rangle} $ as the unbiased estimator of Eq. (\ref{Eq:vqd}).
    
\end{enumerate}

Note that the sign $h_{l_k}$ can be considered in the postprocessing step in step 1 (c).  This procedure can be regarded as a generalization of the VQED method~\cite{PhysRevA.108.042426}: while VQED uniformly samples the stabilizer operators because the projector onto the symmetric subspace is a linear combination of stabilizer operators with equal weight in Pauli-stabilizer codes, our method samples displacement operators as stabilizer operators with the weight of $p_{\vec{l},\vec{l}'}$. One may consider this method as an implementation of LCU with a random sampling of unitary operators~\cite{faehrmann2022randomizing,wan2022randomized,chakraborty2024implementing}. Due to the division, the estimator's variance grows with the squared inverse of the projection probability~\cite{PhysRevA.108.042426}. When we use virtual PS for state preparation, the number of samples to certify the accuracy $\varepsilon$ scales as $N_{\rm s}=O(C_{z, \delta z}^{N_{\rm q}} \varepsilon^{-2})$ and $N_{\rm s}=O(C_{\Delta, s}^{N_{\rm q}} \varepsilon^{-2})$ for the SC states and the rectangular GKP states, with $C_{z, \delta z}= q_{z, \delta z}^{-2}$ and $C_{\Delta, s}= q_{\Delta, s}^{-2}$. We remark that the VQED for rotation symmetries can also be incorporated for raising the robustness against photon loss~\cite{endo2022quantum}, which is realized with VQED by using the quantum circuit shown in Fig. \ref{Fig:vqsd} (d3).

\textcolor{black}{
Let $T_{\rm VQED}$ denote the required gate times for the VQED implementation except for the measurement process.}
\textcolor{black}{Because we randomly generate the gate, $T_{\rm VQED}$ is also a random variable. Denote the smeared projector as $\tilde{P} \propto \sum_x e^{-x^2/ \sigma^2} D(x)$. By assuming the gate time is proportional to the displacement amount $x$, the most frequently generated $T_{\rm VQED}$ is proportional to $\sigma$ since $\mathrm{Var}[x] \propto \sigma^2$. Then, $\sigma=\Gamma$ and $\sigma=\Gamma_0$ for SC and the square GKP codes result in the typical gate time $T_{\rm VQED} = O(e^{z} \sqrt{e^{2 \delta z} -1})$, which is quadratically better scaling than the LCU implementation. Note that $T_{\rm VQED}$ is constant if the gate time can be made independent of the displacement amount.}

\textcolor{black}{We now consider cases where $T_1$ and $T_2$ ancilla qubit errors are present and describe how to mitigate their effects. In SM, we show that ancilla qubit errors shrink the simulated process, i.e., $\tilde{\mathcal{P}}_{l_k, l'_k}(\cdot)$ transforms into $e_{l_k, l'_k} \tilde{\mathcal{P}}_{l_k, l'_k}(\cdot)$, where $0 < e_{l_k, l'_k} < 1$. To compensate for this error, based on the characterization of $e_{l_k, l'_k}$, we modify the sampling probability of the indices $(l_k, l'_k)$ to $p'_{l_k, l'_k} = p_{l_k} p_{l'_k} e_{l_k, l'_k}^{-1} / R$, where $R = \sum_{l_k, l'_k} p_{l_k} p_{l'_k} e_{l_k, l'_k}^{-1} > 1$. This adjustment ensures that the numerator and denominator in Eq.~\eqref{Eq:vqd} are equally rescaled, allowing us to obtain an unbiased estimator for $\langle O \rangle_{\rm vs}$ in Eq.~\eqref{Eq:vqd}. The modified sampling introduces a multiplicative sampling overhead of $O(q^{-2}R^{-2})$ per VQED operation, where $q$ is the projection probability in the noiseless case. See SM for detailed derivations. Note that this method can be applied to methods that also simulate a target process by randomly generating Hadamard test circuits and post-processing of measurement outcomes~\cite{sun2022perturbative,chakraborty2024implementing,huo2023error,zeng2021universal,wan2022randomized}. }

\begin{figure}[t!]
    \centering
    \includegraphics[width=\columnwidth]{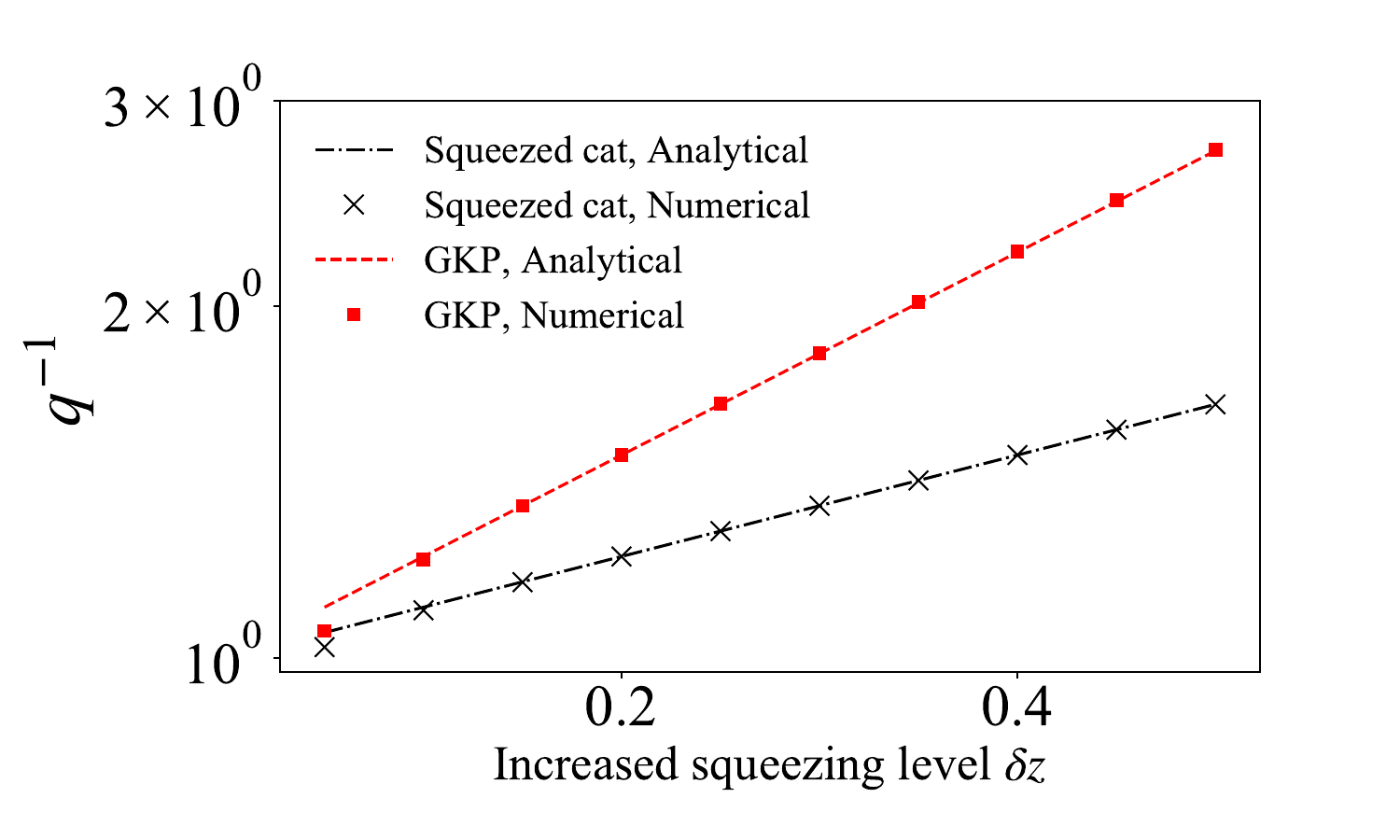}
    \caption{The inverse of the projection probability $q^{-1}$ for the magic state of a square lattice GKP and SC states according to the increased squeezing level $\delta z$. We set the initial GKP state squeezing envelope parameter $\Delta^2=0.05$ and the amplitude and the squeezed level of the SC state to $\xi=\sqrt{\pi/2}$ and $z=-\frac{1}{2}\mathrm{ln}(\Delta^2)$. }
    \label{Fig:Fig2}
\end{figure}

\emph{Numerical simulations.---} We numerically verify the performance of our PS protocol. We simulate the pure magic states $\ket{A}_L \propto \ket{0}_L+e^{i \frac{\pi}{4}}\ket{1}_L$ for GKP and SC codes. We plot the sampling cost of virtual PS in accordance with the virtually squeezed level $\delta z$ in Fig. \ref{Fig:Fig2}. As discussed before, we find that the analytical and numerical results deviate at the small increased squeezing value but converge to the analytical result. We also verify that the expectation values for the logical operators for SC and GKP states have the same trend. See SM for details.

\emph{Conclusion and Discussions.---} In this paper, we propose new squeezing methods, i.e., LCU and virtual PS, by constructing the projector onto the subspace with a higher squeezing level for more accurate computation on translation symmetric bosonic codes. We find that the sampling overhead of the projection for the GKP code subspace is quadratically higher than that of the SC code subspace because squeezing for two quadratures is required for GKP states. 


While both LCU and virtual PS methods can be used for improving computation accuracy in quantum computation, we mention another application of virtual PS. Because the classical simulation of concatenated error correction codes, i.e., the surface code consisting of GKP code states as the physical qubit layer, is classically inefficient for the intermediate squeezing level, virtual PS is useful for evaluating the threshold value or threshold squeezing levels for simulating higher squeeing-level states by using the available quantum states in the actual quantum hardware.

Our PS methods, together with symmetry expansion with rotation symmetries, may be applicable to a broad range of symmetries. For example, they can be straightforwardly leveraged for error suppression of spin-ensemble error correction codes such as spin-squeezed GKP~\cite{omanakuttan2023spin}, spin cat, and spin binomial codes~\cite{albert2018performance} with significantly moderate hardware requirements because the projector onto the code subspace can be constructed similarly with the symmetry operators. Also, it may be interesting to apply our PS methods to recently proposed molecular codes~\cite{albert2020robust}.

\section*{Acknowledgments}
PRESTO, JST supports this work, Grant No. JPMJPR2114; CREST, JST, Grant No.\,JPMJCR1771 and JPMJCR23I4; MEXT Q-LEAP Grant No.\,JPMXS0120319794 and JPMXS0118068682. This work also was supported by JST [Moonshot R\&D][Grant Number JPMJMS2061]. S.E. acknowledges useful discussions with Kosuke Mizuno, Takaaki Takenakada, Rui Asaoka, and Hayata Yamasaki. This work was supported by JST Moonshot (Grant Number JPMJMS226C).
Y. Matsuzaki is supported by JSPS KAKENHI (Grant Number
23H04390) and CREST(JPMJCR23I5).

\bibliographystyle{apsrev4-1}
\bibliography{bib}

\onecolumngrid

\clearpage

\begin{center}
	\Large
	\textbf{Supplementary Materials: Projective squeezing for translation symmetric bosonic codes}
\end{center}

\section{Logical Pauli operators for SC codes}
First, $Z_{\rm{sq}}\equiv e^{i \pi a^\dag a}$ trivially acts as a logical Z operator. Here, we show that $X_{\rm{sq}} \equiv -i D(i \frac{\pi}{4\xi})$ operates as a logical $X$ operator in the SC code. For the projector onto the code space $P_{\rm{sq}}=\ket{\xi,z}\bra{\xi,z}+\ket{-\xi,z}\bra{-\xi,z}$,  we obtain
\begin{equation}
P_{\rm{sq}} X_{\rm{sq}} P_{\rm{sq}} \sim e^{-\frac{1}{2} (\frac{\pi}{4\xi})^2 e^{-2 z} } (\ket{\xi,z}\bra{\xi,z}-\ket{-\xi,z}\bra{-\xi,z}). 
\end{equation}
for sufficiently large $\xi$ and $z$ by using 

\begin{equation}
\bra{\xi,z} D(\eta) \ket{\xi, z} = e^{i 2 \xi y} \mathrm{exp}\big(-\frac{x^2 e^{2z} +y^2 e^{-2z}}{2} \big),
\label{Eq: overlapsq}
\end{equation}
for $\eta=x+iy~(x, y \in \mathbb{R})$. This indicates that $X_{\rm{sq}}$ behaves better as a logical $X$ operator as $\xi$ and $z$ increase on the code space of SC codes since $\ket{\pm \xi,z}$ is a logical $\ket{\pm}$ state.

\section{Projective squeezing for SC states}
Here, we consider the projector for squeezing as follows: 

\begin{equation}
\begin{aligned}
\tilde{P}_{\xi, \Gamma} &= \sum_l p_l^{\xi, \Gamma} (-1)^l D\big(i \frac{\pi}{2 \xi} l\big) \\
p_l^{\xi, \Gamma}& \propto e^{-\frac{1}{\Gamma^2} (\frac{\pi}{2 \xi} l)^2}, ~\sum_l p_l^{\xi, \Gamma}=1.
\label{Eq: projectorapp1}
\end{aligned}
\end{equation}

Then, we have
\begin{equation}
\begin{aligned}
\tilde{P}_{\xi, \Gamma} \ket{\xi, z} &= \sum_l p_{l}^{\xi, \Gamma} (-1)^l D\bigg(i \frac{\pi}{2 \xi} l \bigg)  \ket{\xi, z} \\
& \propto \sum_l p_l^{\xi, \Gamma} (-1)^l \int dp e^{-i \sqrt{2} \xi p} e^{-\frac{1}{2}(p e^{-z})^2} \ket{p+\frac{\pi}{\sqrt{2}\xi} l}_p \\
&= \int dp e^{-i \sqrt{2} \xi p} \big[\sum_l p_l^{\xi, \Gamma} e^{-\frac{1}{2}[(p-\frac{\pi}{\sqrt{2}\xi}l) e^{-z}]^2} \big] \ket{p}_p. 
\end{aligned}
\end{equation}
Here, $\ket{p}_p$ is the eigenvector of the momentum operator $p=\frac{a-a^\dag}{\sqrt{2} i}$.

Now, 

\begin{equation}
\begin{aligned}
\sum_l p_l^{\xi, \Gamma} e^{-\frac{1}{2}[(p-\frac{\pi}{\sqrt{2} \xi}l) e^{-z}]^2} &\propto \sum_l e^{-\frac{1}{\Gamma^2} (\frac{\pi}{2 \xi} l)^2} e^{-\frac{1}{2}[(p-\frac{\pi}{\sqrt{2} \xi}l) e^{-z}]^2} \\
&= \sum_l e^{-\frac{1}{\Gamma^2} (\frac{\pi}{2 \xi})^2 l^2} e^{-e^{-2z} (\frac{\pi}{2\xi})^2(l-\frac{\sqrt{2} \xi}{ \pi} p)^2 }
\label{Eq: sum}
\end{aligned}
\end{equation}

By replacing the summation with the integral, we approximate Eq. (\ref{Eq: sum}) as
\begin{equation}
\begin{aligned}
\sum_l e^{-\frac{1}{\Gamma^2} (\frac{\pi}{2 \xi})^2 l^2} e^{-e^{-2z} (\frac{\pi}{2\xi})^2(l-\frac{\sqrt{2} \xi}{ \pi} p)^2 } &\sim \int dx  e^{-\frac{1}{\Gamma^2} (\frac{\pi}{2\xi})^2 x^2} e^{-e^{-2z} (\frac{\pi}{2\xi})^2 (x-\frac{\sqrt{2}\xi}{\pi} p)^2 } \\
& \propto e^{-\frac{1}{2}\frac{e^{-2z}}{1+\Gamma^2 e^{-2z}} p^2}. 
\label{Eq: replace1}
\end{aligned}
\end{equation}
Here, we use $\int dx e^{-a x^2} e^{-b(x-c)^2} =\frac{ \sqrt{\pi} e^{-\frac{ abc^2}{a+b}}}{\sqrt{a+b}}$ for $a,b >0$ and $c \in \mathbb{R}$. Thus, for raising the squeezing level to $z+\delta z$ from $z$, we need
\begin{equation}
e^{-2 \delta z} = \frac{1}{1+\Gamma^2 e^{-2 z}}, 
\label{Eq: sqandgamma}
\end{equation}
and we have
\begin{equation}
\tilde{P}_{\xi, \Gamma} \ket{\xi, z} \sim \ket{\xi, z+\delta z}. 
\end{equation}

Now, we discuss the projection probability for PS. The projection probability can be described as
\begin{equation}
q_{\delta z} = \bra{\xi, z} \tilde{P}_{\xi, \Gamma}^2 \ket{\xi, z},
\end{equation}
where
\begin{equation}
\tilde{P}_{\xi, \Gamma}^2 = \frac{\sum_{l l'} e^{-\frac{1}{\Gamma^2} (\frac{\pi}{2 \xi})^2 (l^2+l'^2)} D(i \frac{\pi}{2 \xi} (l+l')) (-1)^{l+l'}}{\sum_{l l'} e^{-\frac{1}{\Gamma^2} (\frac{\pi}{2 \xi})^2 (l^2+l'^2)}}. 
\end{equation}
Then, we obtain via Eq. (\ref{Eq: overlapsq})
\begin{equation}
q_{\delta z}= \frac{\sum_{l l'} e^{-\frac{1}{\Gamma^2} (\frac{\pi}{2 \xi})^2 (l^2+l'^2)} e^{-\frac{e^{-2z}}{2} (\frac{\pi}{2\xi})^2(l+l')^2}}{\sum_{l l'} e^{-\frac{1}{\Gamma^2} (\frac{\pi}{2 \xi})^2 (l^2+l'^2)}}. 
\label{Eq: qdeltaz}
\end{equation}
By replacing the summation with the integral, we obtain 
\begin{equation}
\begin{aligned}
\sum_{l l'} e^{-\frac{1}{\Gamma^2} (\frac{\pi}{2 \xi})^2 (l^2+l'^2)} e^{-\frac{e^{-2z}}{2} (\frac{\pi}{2\xi})^2(l+l')^2}&\sim \int dx dx' e^{-\frac{1}{\Gamma^2} (\frac{\pi}{2 \xi})^2 (x^2+x'^2)} e^{-\frac{e^{-2z}}{2} (\frac{\pi}{2\xi})^2(x+x')^2} \\
&= \frac{\pi \Gamma}{\big(\frac{\pi}{2\xi}\big)^2 \sqrt{\frac{1}{\Gamma^2}+e^{-2z}} }, 
\label{Eq: replace2}
\end{aligned}
\end{equation}
and 
\begin{equation}
\begin{aligned}
\sum_{l l^\prime} e^{-\frac{1}{\Gamma^2 } (\frac{\pi}{2 \xi})^2 (l^2+l'^2)} &\sim \int dx dx' e^{-\frac{1}{\Gamma^2 } (\frac{\pi}{2 \xi})^2 (x^2+x'^2)}  \\
&=\frac{\pi \Gamma^2}{(\frac{\pi}{2\xi})^2}.
\label{Eq: replace3}
\end{aligned}
\end{equation}
Here, we use $\int dx dx' e^{-a(x^2+x'^2)} e^{-b(x+x')^2} =\frac{\pi}{\sqrt{a(a+2b)}}$. Thus, we obtain 
\begin{equation}
\begin{aligned}
q_{\delta z}&\sim \frac{1}{\sqrt{1+\Gamma^2 e^{-2 z}}} \\
&= e^{-\delta z},
\end{aligned}
\end{equation}
where we use Eq. \eqref{Eq: sqandgamma}. This yields the sampling overhead of virtual PS $C_{\delta z}=q_{\delta z}^{-2} = e^{2 \delta z}$. 

Then, the SC code states can be written as:

\begin{equation}
\ket{\mathrm{sq}_{\xi,z}^{\mu} }=\frac{1}{N_{\rm sq}} (\ket{\xi, z} \pm \ket{-\xi, z}), 
\end{equation}
where $N_{\rm sq}$ is the normalization factor. Regarding the approximately orthogonal squeezed coherent states, $N_{\rm sq} \sim \sqrt{2}$. In the same procedure above, we can show $\tilde{P}_{\xi, \Gamma} \ket{-\xi, z}\sim \ket{-\xi, z+\delta z}$ for $\Gamma$ satisfying Eq. \eqref{Eq: sqandgamma}. Therefore, we have
\begin{equation}
\tilde{P}_{\xi, \Gamma} \ket{\mathrm{sq}_{\xi,z}^{\mu} } \sim  \ket{\mathrm{sq}_{\xi,z+\delta z}^{\mu} }. 
\end{equation}
Next, the projection probability reads:
\begin{equation}
\begin{aligned}
q_{\delta z}&= \bra{\mathrm{sq}_{\xi,z+\delta z}^{\mu} } \tilde{P}_{\xi, \Gamma}^2 \ket{\mathrm{sq}_{\xi,z+\delta z}^{\mu} } \\
&\sim \frac{1}{2} (\bra{\xi, z} \tilde{P}_{\xi, \Gamma}^2 \ket{\xi, z}+\bra{-\xi, z} \tilde{P}_{\xi, \Gamma}^2 \ket{-\xi, z}) \\
&=\bra{\xi, z} \tilde{P}_{\xi, \Gamma}^2 \ket{\xi, z}=e^{-\delta z}, 
\label{Eq: sqinvprob}
\end{aligned}
\end{equation}
where we ignore the non-diagonal terms by assuming the overlap between $\ket{\xi, z}$ and $\ket{-\xi, z}$ is negligible. The sampling overhead for virtual PS again reads $C_{\delta z}=e^{2 \delta z}$. 

\begin{figure}[t!]
    \centering
    \includegraphics[width=0.7\columnwidth]{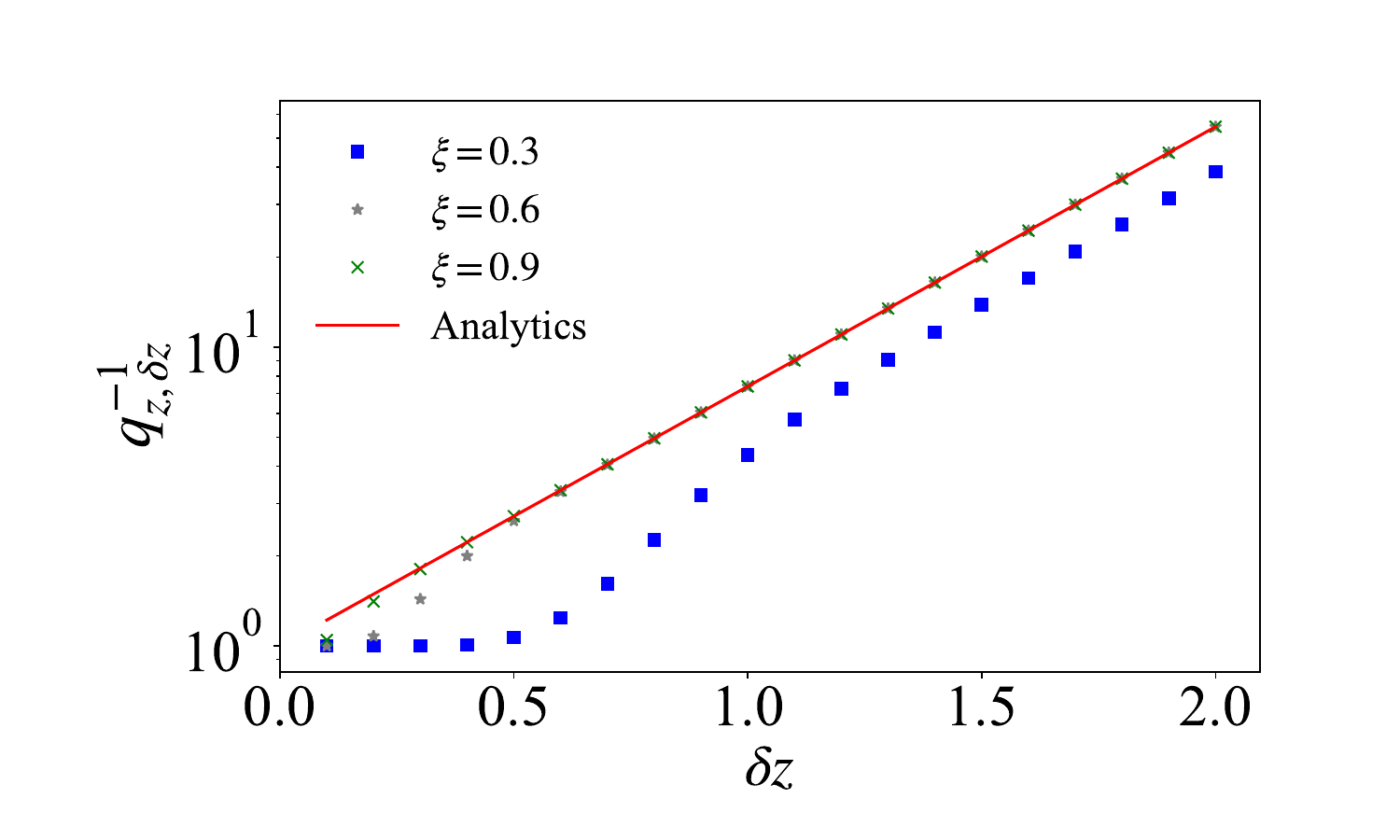}
    \caption{The inverse projection probability $q_{z, \delta z}^{-1}$ versus the increased squeezing level $\delta z$ for $z=-\mathrm{log} (0.3)$. We plot the analytical inverse probability $q_{z, \delta z}^{-1} = e^{\delta z}$ and numerically calculated ones from Eq. (\ref{Eq: qdeltaz}) for $\xi=0.3, 0.6, 0.9$.  }
    \label{Fig: sqvsprob}
\end{figure}

Note that the replacement of summation in Eqs. (\ref{Eq: replace1}, \ref{Eq: replace2}, \ref{Eq: replace3}) with the integral is justified for the following conditions:
\begin{equation}
e^{2 z} \gtrsim \bigg(\frac{\pi}{2 \xi} \bigg)^2
\label{Eq: Cond1}
\end{equation}
and 
\begin{equation}
\delta z \gtrsim \frac{1}{2} e^{-2 z} \bigg(\frac{\pi}{2 \xi} \bigg)^2
\label{Eq: Cond2}
\end{equation}
because the summation interval is sufficiently smaller than the width of the Gaussian in this regime. In Fig. \ref{Fig: sqvsprob}, we plot the inverse of the projection probabilities of Eq. (\ref{Eq: qdeltaz}) versus the increased squeezing value $\delta z$ for different amplitudes $\xi$. We also plot the analytical approximation in Eq. (\ref{Eq: sqinvprob}). Note that the first condition (\ref{Eq: Cond1}) is not satisfied for $\xi=0.3$, and therefore, the inverse of the projection probability does not converge to the analytical one. Meanwhile, $\xi=0.6, 0.9$ meets the first condition. In this case, as $\delta z$ increases, the second condition (\ref{Eq: Cond2}) is satisfied, resulting in the convergence to the analytical approximation. Note that $\xi=0.9$ shows faster convergence than $\xi=0.6$ as the condition (\ref{Eq: Cond2}) indicates. 

\section{Projective squeezing for GKP code states}
Here, we consider the following projector:

\begin{equation}
\begin{aligned}
\tilde{P}_{\xi, \vec{\Gamma}}&=\sum_{\vec{l}} p_{\vec{l}}^{\xi, \vec{\Gamma}} D\left(2 \alpha l_1 +2 \beta l_2\right), \\
p_{\vec{l}}^{\xi, \vec{\Gamma}} &\propto e^{-\frac{1}{{\Gamma_1}^2} (2 |\alpha| l_1)^2} e^{-\frac{1}{{\Gamma_2}^2} (2|\beta| l_2)^2}, \sum_{\vec{l}} p_{\vec{l}}^{\xi, \vec{\Gamma}} =1
\label{Eq: projector2}
\end{aligned}
\end{equation}
with $\alpha=\xi$, $\beta=i \frac{\pi}{2\xi}$ for $\xi>0$,  $\vec{l}=(l_1, l_2) \in \mathbb{Z}^2$ and $\vec{\Gamma}=(\Gamma_1, \Gamma_2) \in \mathbb{R}^2$. The approximate GKP states reads as 
\begin{equation}
\ket{\rm{gkp}_\mu^{\Delta}} \propto e^{-\Delta^2 a^\dag a} \ket{\rm{gkp}_\mu}, 
\end{equation}
for the ideal GKP code states $\ket{\rm{gkp}_\mu} ~(\mu=0,1)$. Note that we have 
\begin{equation}
\begin{aligned}
e^{-\Delta^2 a^\dag a}&= \frac{1}{\pi(1-e^{-\Delta^2})} \int dX dY e^{-\frac{X^2+Y^2}{2 \mathrm{tanh}(\Delta^2/2)}} D(X+iY) \\
&\sim \int dX dY \frac{e^{-(X^2+Y^2)/\Delta^2}}{\pi \Delta^2}D(X+iY), 
\end{aligned}
\end{equation}
with $X, Y \in \mathbb{R}$ for $\Delta^2 \ll 1$~\cite{royer2020stabilization}.

Now, we apply the projector $\tilde{P}_{\xi, \vec{\Gamma}}$ to the state $\ket{\rm{gkp}_\mu^{\Delta}}$: 
\begin{equation}
\begin{aligned}
\tilde{P}_{\xi, \vec{\Gamma}} \ket{\rm{gkp}_\mu^{\Delta}} &\propto \tilde{P}_{\xi, \vec{\Gamma}}  e^{-\Delta^2 a^\dag a}\ket{\rm{gkp}_\mu} \\
&\propto \sum_{\vec{l}} p_{\vec{l}}^{\xi, \vec{\Gamma}} \int dX dY e^{-\frac{X^2+Y^2}{2 \mathrm{tanh}(\Delta^2/2)}}  D\left(2 \alpha l_1 +2 \beta l_2\right) D(X+iY) \ket{\rm{gkp}_\mu} \\
&=  \int dX dY e^{-\frac{X^2+Y^2}{2 \mathrm{tanh}(\Delta^2/2)}}  \left(\sum_{\vec{l}} p_{\vec{l}}^{\xi, \vec{\Gamma}}e^{-i 4 |\alpha| Y l_1} e^{-i 4 |\beta| Xl_2} \right) D (X+ iY) \ket{\rm{gkp}_\mu}, 
\label{Eq: applicationofprojector}
\end{aligned}
\end{equation}
where we use $D\left(2 \alpha l_1 +2 \beta l_2\right) \ket{\rm{gkp}_\mu}=\ket{\rm{gkp}_\mu}$ and $D(\gamma_1) D(\gamma_2)=e^{\gamma_1 \gamma_2^* -\gamma_1^* \gamma_2} D(\gamma_2) D(\gamma_1)$ for $\gamma_1, \gamma_2 \in \mathbb{C}$ in the third line. Then, we have
\begin{equation}
\sum_{\vec{l}} p_{\vec{l}}^{\xi, \vec{\Gamma}}e^{-i 4 |\alpha| Y l_1} e^{-i 4 |\beta| X l_2} = \left(\sum_{l_1}  e^{-\frac{1}{{4 \Gamma_1}^2} (4 |\alpha| l_1)^2} e^{-i 4 |\alpha| l_1 Y } \right) \left(\sum_{l_2}  e^{-\frac{1}{4{\Gamma_2}^2} (4 |\beta| l_2)^2} e^{-i 4 |\beta| l_2 X}\right). 
\end{equation}

The term $\sum_{l_1}  e^{-\frac{1}{{4 \Gamma_1}^2} (4 |\alpha| l_1)^2} e^{-i 4 |\alpha| l_1 Y }$ corresponds to the discrete Fourier transform, which becomes a periodic function of $\frac{\pi}{2|\alpha|}$ as a function of $Y$. For $|Y| \leq \pi/ 2|\alpha|$, this function is well-approximated by $e^{-\Gamma_1^2 Y^2} $ except for a multiplicative constant. The same arguments hold for $\sum_{l_2}  e^{-\frac{1}{4{\Gamma_2}^2} (4 |\beta| l_2)^2} e^{-i 4 |\beta| l_2 X}$. Considering the envelope of $e^{-\frac{X^2+Y^2}{2 \mathrm{tanh}(\Delta^2/2)}}$ in Eq. (\ref{Eq: applicationofprojector}), the range of $|X| \leq \pi/ 2|\beta|$ and $|Y| \leq \pi/ 2|\alpha|$ has a dominant effect on the calculation, and therefore we approximate Eq. (\ref{Eq: applicationofprojector}) for $\Gamma_1=\Gamma_2=\Gamma_0$ as:
\begin{equation}
\begin{aligned}
\tilde{P}_{\xi, \vec{\Gamma}} \ket{\rm{gkp}_\mu^{\Delta}} &\sim  \frac{1}{\mathcal{N}}\int dX dY e^{-\left(\frac{1}{\Delta^2}+\Gamma_0^2 \right) (X^2+Y^2)}  D(X+iY) \ket{\mathrm{gkp}_\mu} \\
&\propto e^{-\frac{1}{1/{\Delta^2} +\Gamma_0^2} a^\dag a}   \ket{\mathrm{gkp}_\mu},
\label{Eq: GKPstateapproximation}
\end{aligned}
\end{equation}
where we also approximate $\mathrm{tanh}(\Delta^2/2) \sim \Delta^2/2$ for sufficiently small $\Delta^2$, and $\mathcal{N}$ is the normalization factor. Therefore, by setting $\Gamma_0^2=\frac{1}{\Delta^2}(s-1)$, we can virtually simulate the GKP states for $\Delta'^2 = \frac{\Delta^2}{s}$.

We now discuss the sampling cost for PS for GKP states. Here, we consider the square lattice GKP code case, i.e., $|\alpha|=|\beta|=\sqrt{\pi/2}$. The projection probability reads: 
\begin{equation}
q_{\Delta,s}= \Tr[\tilde{P}_{\xi, \vec{\Gamma}}^\dag \tilde{P}_{\xi, \vec{\Gamma}} \ket{\mathrm{gkp}_\mu^{\Delta}}\bra{\mathrm{gkp}_\mu^{\Delta}}]. 
\end{equation}
Since $\tilde{P}_{\xi, \vec{\Gamma}}^\dag=\tilde{P}_{\xi, \vec{\Gamma}}$, $p_{\vec{l}}^{\xi, \vec{\Gamma}} \propto e^{-\frac{1}{{\Gamma}^2} (2 |\alpha| l)^2} e^{-\frac{1}{{\Gamma}^2} (2 |\beta| l_2)^2}$ and $\sum_{\vec{l}} p_{\vec{l}}^{\xi, \vec{\Gamma}} =1$, we obtain 
\begin{equation}
\begin{aligned}
\tilde{P}_{\xi, \vec{\Gamma}}^\dag \tilde{P}_{\xi, \vec{\Gamma}}&= \frac{1}{\mathcal{M}^2} \sum_{l_1, l_2, l_1', l_2'} e^{-\frac{l_1^2+l_1^{\prime 2}+l_2^2+l_2^{\prime 2}}{ \Gamma_0^2}(2|\alpha|)^2 } D(2\alpha (l_1+l_1')+2\beta (l_2+l_2')) \\
\mathcal{M}&= \sum_{l_1, l_2} e^{-\frac{l_1^2+l_2^2}{\Gamma_0^2} (2|\alpha|)^2}.
\end{aligned}
\end{equation}
Then, we have
\begin{equation}
q_{\Delta, s}= \frac{1}{\mathcal{M}^2} \sum_{l_1, l_2, l_1', l_2'} e^{-\frac{l_1^2+l_1^{\prime 2}+l_2^2+l_2^{\prime 2}}{ \Gamma_0^2} (2|\alpha|)^2 } \bra{\mathrm{gkp}_\mu^{\Delta}} D(2\alpha (l_1+l_1')+2\beta (l_2+l_2'))  \ket{\mathrm{gkp}_\mu^{\Delta}}.
\label{Eq: qdeltas}
\end{equation}

To further calculate Eq. (\ref{Eq: qdeltas}), we use 
\begin{equation}
\bra{\mathrm{gkp}_\mu^{\Delta}} D(\lambda) \ket{\mathrm{gkp}_\mu^{\Delta}} \sim \sum_{\vec{k} \in \mathbb{Z}^2} e^{i \pi (k_1 +\mu)k_2} e^{-\frac{1}{2 \Delta^2} |\lambda-\Lambda_0^{\vec{k}}|^2} e^{-\frac{\Delta^2}{2} |\Lambda_0^{\vec{k}}|^2},
\label{Eq: overlap}
\end{equation}
where $\lambda \in \mathbb{C}$, $\Lambda_0^{\vec{k}}= |\alpha| (2 k_1 +i k_2)$~\cite{albert2018performance}. For more general results, including the non-diagonal terms, refer to Eq. (7.14) in Ref. \cite{albert2018performance}. For $\Delta \rightarrow 0$, $e^{-\frac{1}{2 \Delta^2} |\lambda-\Lambda_0^{\vec{k}}|^2} \sim \delta_{\lambda, \Lambda_0^{\vec{k}}}$, where $\delta_{x,y}$ is Kronecker delta. Therefore, for $\lambda= 2|\alpha|(n_1+i n_2),~ n_1, n_2 \in \mathbb{Z}$, Eq. (\ref{Eq: overlap}) can be further approximated as:
\begin{equation}
\bra{\mathrm{gkp}_\mu^{\Delta}} D(\lambda) \ket{\mathrm{gkp}_\mu^{\Delta}} \sim e^{-2\Delta^2 |\alpha|^2 (n_1^2+n_2^2)}.
\label{Eq: overlap2}
\end{equation}

These approximations become increasingly accurate for small $\Delta$~\cite{albert2018performance}. Then, by substituting Eq. (\ref{Eq: overlap2}) in Eq. (\ref{Eq: qdeltas}), we obtain 
\begin{equation}
\begin{aligned}
q_{\Delta,s} &\sim \frac{1}{\mathcal{M}^2} \sum_{l_1, l_2, l_1', l_2'} e^{-\frac{l_1^2+l_1^{\prime 2}+l_2^2+l_2^{\prime 2}}{\Gamma_0^2} (2|\alpha|)^2 } \times e^{-2 |\alpha|^2 \Delta^2 ((l_1+l_1')^2 + (l_2+l_2')^2)} \\
&= \left( \frac{1}{\mathcal{M}} \sum_{l_1, l_1'} e^{-\frac{l_1^2+l_1'^2}{\Gamma_0^2} (2|\alpha|)^2} e^{-2 \Delta^2 |\alpha|^2 (l_1+l_1')^2} \right)^2.
\label{Eq: probabilityGKP}
\end{aligned}
\end{equation}
Now, we approximate $\mathcal{M} \sim \int dx dy e^{-\frac{x^2+y^2}{\Gamma_0^2} }=\Gamma_0^2 \pi$. Also, 

\begin{equation}
\begin{aligned}
\sum_{l_1, l_1'} e^{-\frac{l_1^2+l_1'^2}{4 \Gamma_0^2} |\alpha|^2} e^{-2 \Delta^2 |\alpha|^2 (l_1+l_1')^2} &\sim \int dx dy e^{-\frac{x^2 +y^2}{\Gamma_0^2}} \times e^{-\frac{\Delta^2}{2} (x+y)^2} \\&=\frac{\Gamma_0^2 \pi}{\sqrt{1+\Gamma_0^2 \Delta^2}}
\label{Eq: GKPcostapproximation}
\end{aligned}
\end{equation}

Therefore, we obtain
\begin{equation}
q_{\Delta, s} \sim \frac{1}{1+ \Gamma_0^2 \Delta^2} = \frac{1}{s} 
\end{equation}
for $\Gamma_0^2= \frac{1}{\Delta^2} (s-1)$.  Also, we need the sampling overhead of $C_{\Delta, s}=q_{\Delta,s}^{-2} =s^2$ for virtual PS.

\begin{figure}[t!]
    \centering
    \includegraphics[width=0.7\columnwidth]{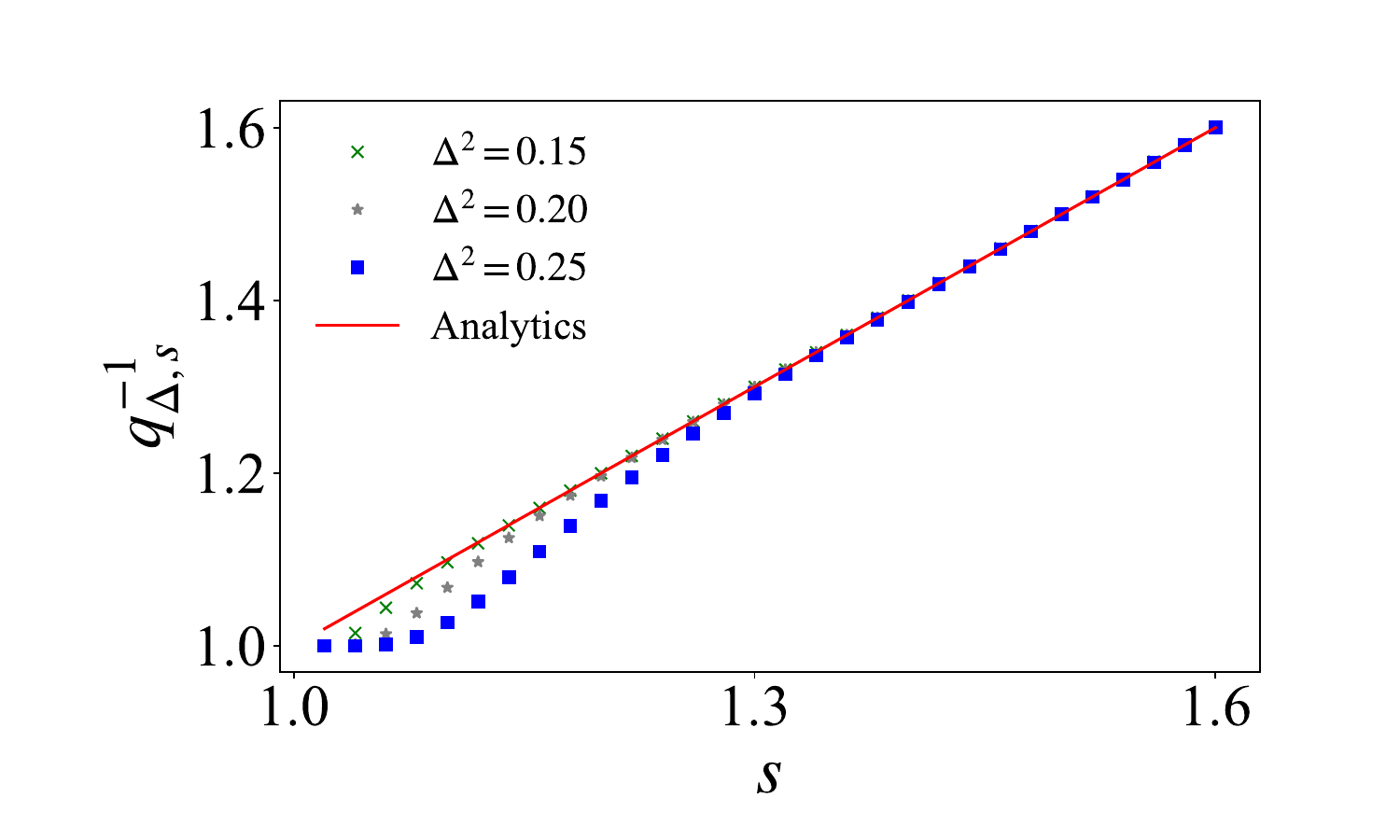}
    \caption{The inverse of the projection probability $q_{\Delta, s}^{-1}$ depending on $s$ for projecting the GKP states for the squeezing parameter $\Delta^2$ onto the logical manifold corresponding to the squeezing parameter $\Delta^2/s$. We plot the analytical inverse probability $q_{\Delta, s}^{-1}=s$ and numerical ones for $\Delta^2=0.15, 0.20, 0.25$.  }
    \label{Fig: GKPprojectionprob}
\end{figure}

Note that the replacement with the integral in Eqs. (\ref{Eq: GKPstateapproximation}, \ref{Eq: GKPcostapproximation}) gives a good approximation when 
\begin{equation}
s \gtrsim 1+ 2 \xi \Delta^2
\label{Eq: Cond3}
\end{equation}
in a similar argument for the SC code. We plot the inverse of the projection probability computed from Eq. (\ref{Eq: probabilityGKP}) versus the value of $s$ in Fig. \ref{Fig: GKPprojectionprob}. We can clearly see that the convergence to the analytical approximation is faster for the small squeezing parameter $\Delta^2$ as the condition (\ref{Eq: Cond3}) indicates. 

\section{Projective squeezing for squeezed vacuum states}
Here, we consider the PS for the squeezed vacuum state. Let the squeezed vacuum state be denoted as $\ket{\xi=0,z}$. Because the infinitely squeezed vacuum state has a translation symmetry toward the momentum direction for an arbitrary value of translation in the case of for $z \in \mathbb{R}$, we can construct the projector as $P_{inf}= \int dz' D(z')$ for $z' \in  \mathbb{R}$. However, since the projection probability onto the infinitely squeezed state is infinitely small, we instead consider the following projector: 
\begin{equation}
\tilde{P}_\gamma \propto \int_{-\infty}^{\infty} dz'  \mathrm{exp}\left( -\frac{z'^2}{\gamma^2} \right) D(z'). 
\label{Eq: vsvacuum}
\end{equation}
Then, we can show that when we apply this projector to a squeezed vacuum state denoted as $\rho_z$ with the squeezing level $z$, $\tilde{P}_{\gamma} \rho_z \tilde{P}_{\gamma}^\dag/{\Tr[\tilde{P}_{ \gamma} \rho_z \tilde{P}_{\gamma}^\dag]}$ is exactly equal to $\rho_{z+\delta z}$ for $\delta z= \frac{1}{2}~ \mathrm{ln} \left(1+ {\gamma^2}{e^{-2 z}} \right)$ without any approximation with a similar calculation procedure to that of SC states. For applying this projector, because the integral involves an infinite number of displacement operators, virtual PS is preferred for the implementation. The projection probability reads $q_{\delta z}=\Tr[\tilde{P}_{\gamma} \rho_z \tilde{P}_{\gamma}^\dag] = e^{- \delta z}$, and 
\begin{equation}
C_{\delta z} \equiv q_{\delta z}^{-2} = e^{2 \delta z} 
\end{equation}
is the sampling overhead.

\section{Expectation values for logical Pauli operators}
In this section, we investigate the expectation values of logical Pauli operators. For SC codes, the logical Pauli operators are $X_{\rm{sq}}=-iD(i\frac{\pi}{4\xi})$, $Z_{\rm{sq}}=e^{i \pi a^\dag a}$, and $Y_{\rm{sq}}=i X_{\rm{sq}} Z_{\rm{sq}}$, as discussed in the previous section. Meanwhile, the logical operators for the rectangular GKP codes are $X_{\rm{GKP}}=D(\xi)$,  $Z_{\rm{GKP}}=D\big( i \frac{\pi}{\xi} \big)$, and $Y_{\rm{GKP}}=i X_{\rm{GKP}} Z_{\rm{GKP}}$.

Here, we discuss the expectation values of logical operators when we apply PS for noiseless states. We compare the outcomes under PS with the analytical results corresponding to the target squeezing level and the exact result for the infinite squeezing level. First, by using Eq. (\ref{Eq: overlapsq}), we can straightforwardly find that the expectation values of the logical X and Y operators for noiseless SC states are the product of $h(\xi, z)= \mathrm{exp}(-\frac{1}{2} (\frac{\pi}{4\xi})^2 e^{-2z})$ and the ideal value, while the logical Z expectation value is not affected by the squeezing level. On the other hand, the expectation values of logical X and Z operators of the GKP states are approximately proportional to $g_1(\Delta, \xi) = \mathrm{exp}(-\frac{\Delta^2 \xi^2}{2})$, with the logical Y expectation values being approximately proportional to $g_2(\Delta, \xi) = \mathrm{exp}(-\Delta^2 \xi^2)$. We compare the expectation values of logical operators under PS with the analytical approximations and the exact ones, depending on the increased squeezing level $\delta z$ for a randomly generated qubit state $\ket{\psi_{\rm rand}}=\ket{\psi'_{\rm rand}}/\|\ket{\psi'_{\rm rand}}\|$, where $\ket{\psi'_{\rm rand}}=c_0 \ket{\tilde{0}} + c_1 \ket{\tilde{1}}$ for $c_0=0.451980+0.329655 i$ and $c_1=0.638855+0.528114i$. We confirm that our PS method can successfully improve the expectation values and show an excellent agreement with the analytical approximations except for small $\delta z$ for both the GKP and the SC state.

\begin{figure}[t!]
    \centering
    \includegraphics[width=0.8\columnwidth]{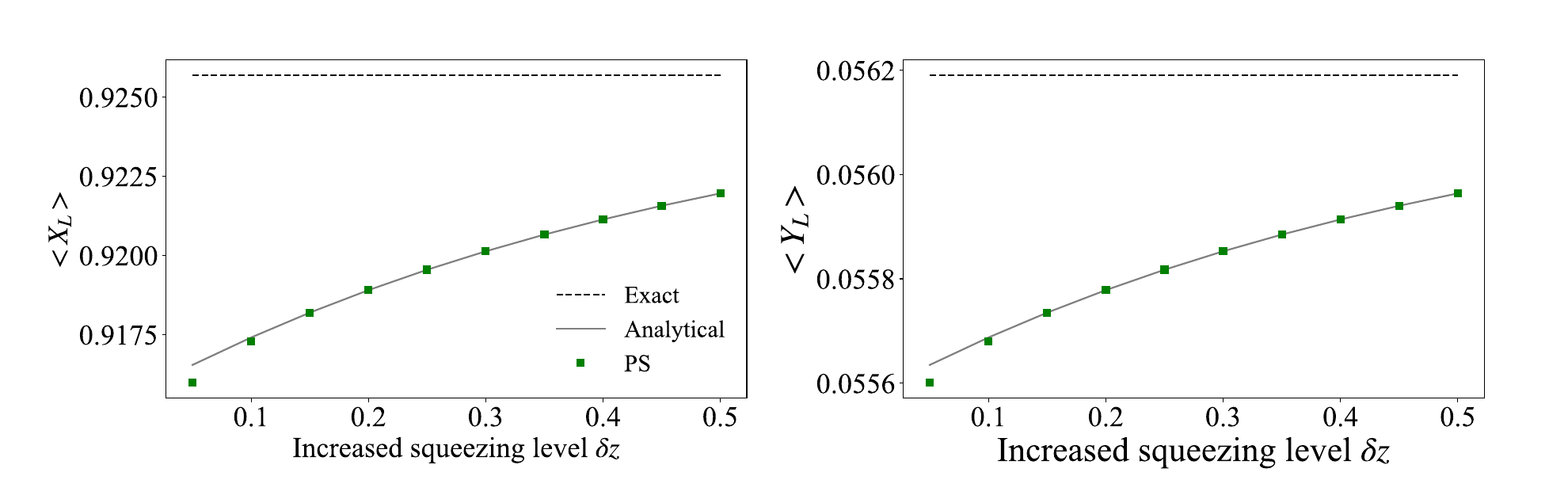}
    \caption{Expectation values of logical Pauli operators for the squeeze cat code state for $z= -\frac{1}{2} \mathrm{log}(0.05)$ under PS according to the increased squeezing level $\delta z$. We plot the exact expectation values for the infinite squeezing level (dashed Black), analytical approximate expectation values for the finite squeezing level (Gray lined), and expectation values under PS (Green square).    }
    \label{Fig: sqcatlogical}
\end{figure}

\begin{figure}[t!]
    \centering
    \includegraphics[width=\columnwidth]{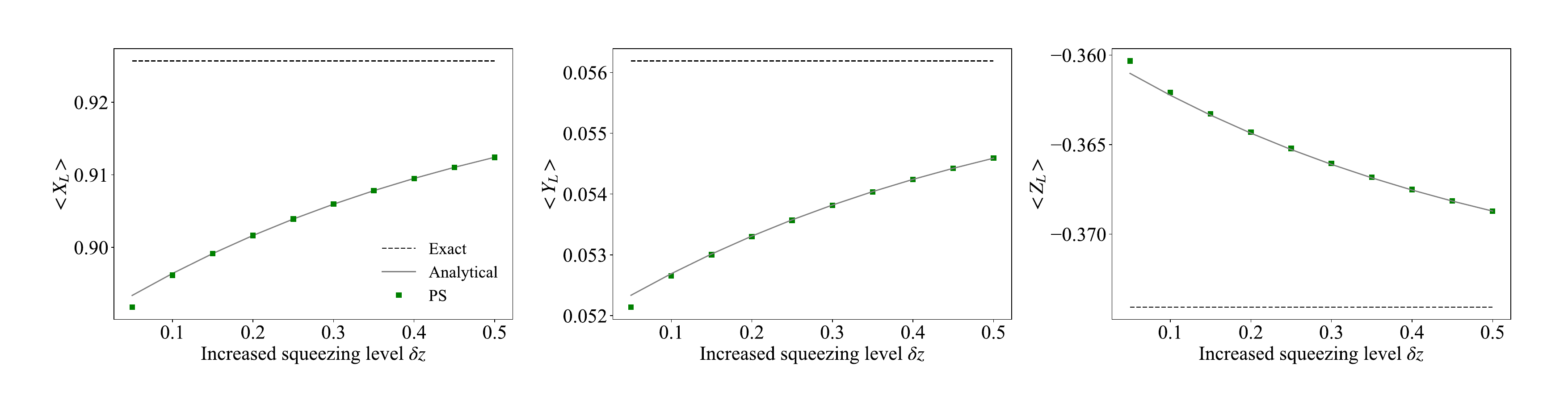}
    \caption{Expectation values of logical Pauli operators for the GKP code state for $\Delta^2=0.05$ under PS according to the increased squeezing level $\delta z$. We plot the exact expectation values for the infinite squeezing level (dashed Black), analytical approximate expectation values for the finite squeezing level (Gray lined), and expectation values under PS (Green square).  }
    \label{Fig: GKPlogical}
\end{figure}

\section{Projective squeezing for noisy states under photon loss}
\begin{figure}[t!]
    \centering
    \includegraphics[width=\columnwidth]{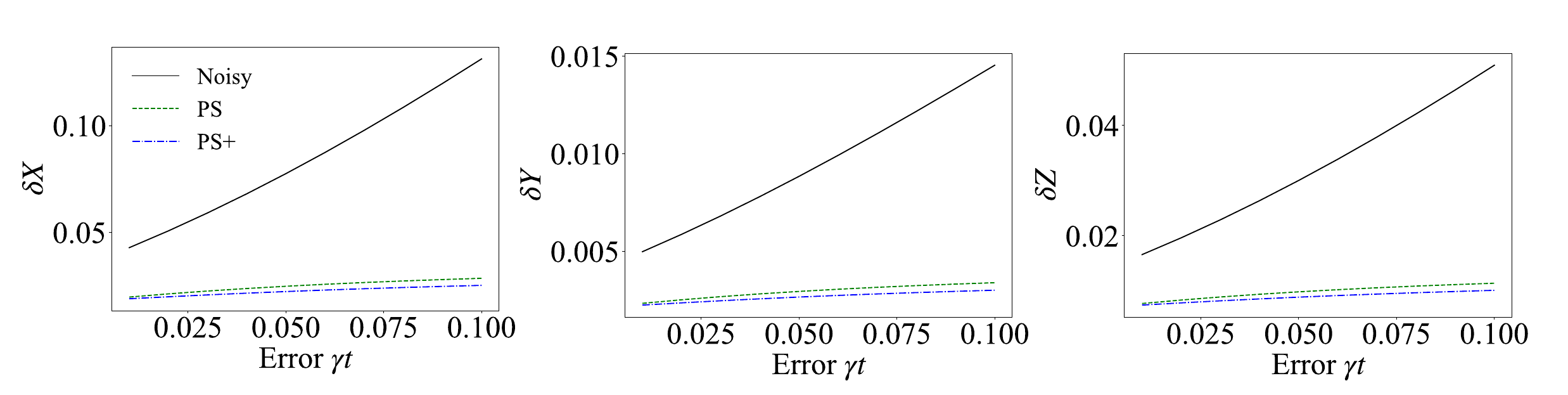}
    \caption{Expectation values of logical Pauli operators for the GKP code state for $\Delta^2=0.05$ under PS for $s=2$ according to the photon loss error $\gamma t$. We plot the absolute value of the difference of expectation values of logical Pauli operators from the ideal one with (Green dotted) and without (Black lined) PS. We also show the result obtained under the symmetry expansion for rotation symmetry followed by PS (Blue dash-dotted).   }
    \label{GKPphotonlossref}
\end{figure}

We also simulate the effect of our PS protocol for the photon loss error $\frac{d \rho}{dt} = \frac{\gamma}{2} (2 a\rho a^\dag -a^\dag a \rho- \rho a^\dag a)$, where $\gamma$ is the photon loss rate. We plot the deviation of expectation values with and without PS for $s=2$ and $\delta z = \frac{1}{2} \mathrm{log} (2)$ for the randomly generated GKP and SC states $\ket{\psi_{\rm rand}}$. We also simulate the case of appending the projection onto the rotation symmetric subspace for the GKP state before applying PS. We find that the GKP state's expectation values show systematic improvement, as shown in Fig. \ref{GKPphotonlossref}, which implies that both the translation and rotation symmetries contribute to the error suppression. For the SC state, an enhancement is apparent in the expectation value of the logical X operator. However, the improvements in the expectation values of the logical Y and Z operators are much more subtle. We leave a detailed analysis of the effect of PS against photon loss for future work. 

\begin{figure}[t!]
    \centering
    \includegraphics[width=\columnwidth]{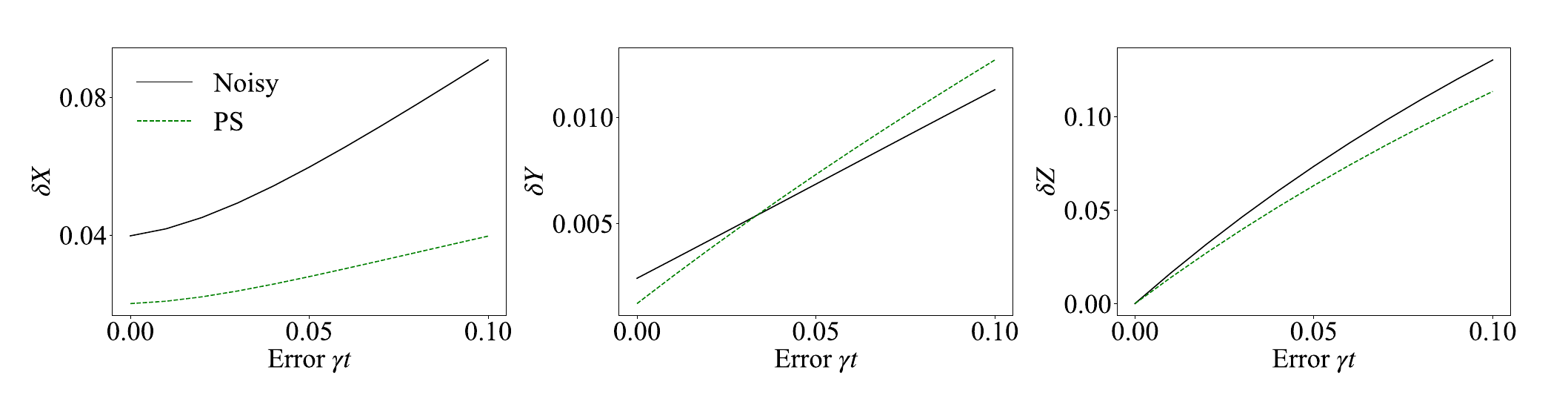}
    \caption{Expectation values of logical Pauli operators for the squeezed code state for $z=-\frac{1}{2} \mathrm{log} (0.05)$ under PS for $s=2$ according to the photon loss error $\gamma t$. We plot the difference of expectation values of logical Pauli operators from the ideal one with (Green dotted) and without (Black lined) PS. We set the increased squeezing level for PS $\delta z =\frac{1}{2} \mathrm{log} (2)$.   }
    \label{Fig2}
\end{figure}

\section{Noise effect in the ancilla qubit}
\textcolor{black}{Here, we first illustrate how the noise effect in the ancilla qubit is described. Then, we discuss the method to mitigate the effect of the ancilla error in the virtual PS implementation. }

We model the noisy dynamics under $T_1$ and $T_2$ errors for the general control operation for the composite system state $\rho_{\rm com} (t)$ by
\begin{equation}
\begin{aligned}
\frac{d}{dt} \rho_{\rm com}(t) = -i [Z \otimes H' ,  \rho_{\rm com}(t)] + \gamma_1 \mathcal{L}[\sigma^{-1} \otimes I]( \rho_{\rm com}(t))+  \gamma_2 \mathcal{L}[Z \otimes I]( \rho_{\rm com}(t)),
\label{Eq: Lindmodel}
\end{aligned}
\end{equation}
where $\mathcal{L}[A](\rho)= \frac{1}{2} (2 A \rho A^\dag -A^\dag A \rho- \rho A^\dag A)$ and the interaction Hamiltonian $Z \otimes H'$ for an Hermite operator $H'$ leads to the target control operation, and $\sigma^{-}=\ket{0}\bra{1}$. Then, we denote the composite state by the block matrix representation 
\begin{equation}
\rho_{\rm com}=\begin{pmatrix}
\rho_{00} & \rho_{01} \\
\rho_{10} & \rho_{11}
\end{pmatrix},
\label{Eq:com}
\end{equation}
where $(i,j)$ corresponding to $\rho_{ij}$ in Eq. \eqref{Eq:com} corresponds to the ancilla qubit states in the computation basis. For example, for the composite state $\ket{+}\bra{+} \otimes \rho_{\rm{r}}$ for a resonator state $\rho_{\rm{r}}$ is represented by
\begin{equation}
\ket{+}\bra{+} \otimes \rho_{\rm{r}} =\frac{1}{2}\begin{pmatrix}
\rho_{\rm{r}} & \rho_{\rm{r}} \\
\rho_{\rm{r}} & \rho_{\rm{r}}
\end{pmatrix}.
\end{equation}

By comparing both sides of Eq. \eqref{Eq: Lindmodel} in the block matrix representation, we obtain 
\begin{equation}
\begin{aligned}
\frac{d \rho_{00}}{dt} &= -i [H', \rho_{00}]+\gamma_1 \rho_{11} \\
\frac{d \rho_{11}}{dt} &= i [H', \rho_{11}]-\gamma_1 \rho_{11} \\
\frac{d \rho_{01}}{dt} &= -i \{H', \rho_{01} \}- \bigg( \frac{\gamma_1}{2} + 2 \gamma_2 \bigg) \rho_{01} \\
\frac{d \rho_{10}}{dt} &= i \{H', \rho_{10} \}- \bigg( \frac{\gamma_1}{2} + 2 \gamma_2 \bigg) \rho_{10}.
\end{aligned}
\end{equation}
For our PS method, only $\rho_{01}$ and $\rho_{10}$ are relevant because we measure the expectation value of the Pauli X operator of the ancilla qubit; therefore, we consider these terms. As solutions for $\rho_{01}$ and $\rho_{10}$, we have 
\begin{equation}
\begin{aligned}
\rho_{01}(t)&= e^{-(\frac{\gamma_1}{2}+\gamma_2) t} \rho_{01}^{\rm id}(t) \\
\rho_{10}(t)&= e^{-(\frac{\gamma_1}{2}+\gamma_2) t} \rho_{10}^{\rm id}(t)
\end{aligned}
\end{equation}
with $\rho_{01}^{\rm id} (t)$ and $\rho_{10}^{\rm id} (t)$ being the solutions for the noiseless case. Note that we finally measure the Pauli $X$ operator in the virtual PS protocol, which yields: 
\begin{equation}
\begin{aligned}
\mathrm{Tr}_{\rm anc}[\rho_{\rm com} (t) X \otimes I] &= \rho_{01} (t) + \rho_{10} (t) \\
&= e^{-(\frac{\gamma_1}{2} + 2\gamma_2) t} (\rho_{01}^{\rm id} (t)+ \rho_{10}^{\rm id} (t) ), \\
\mathrm{Tr}[\rho_{\rm com} (t) X \otimes O] &=\mathrm{Tr}[ (\rho_{01} (t) + \rho_{10} (t)) O] \\
&= e^{-(\frac{\gamma_1}{2} + 2 \gamma_2) t} \mathrm{Tr}[(\rho_{01}^{\rm id} (t)+ \rho_{10}^{\rm id} (t) ) O], 
\label{Eq: uniformdecay}
\end{aligned}
\end{equation}
where $\mathrm{Tr}_{\rm anc}$ denotes the partial trace over the ancilla qubit. Therefore, when we measure the Pauli $X$ expectation value, the computation result only scales with the factor of $e^{-(\frac{\gamma_1}{2}+2 \gamma_2) t}$.

Now, we discuss the robust implementation of the virtual PS against the ancilla qubit noise. We assume we can characterize the scaling factor in advance from the information on ancilla noise errors and the required evolution time for the gate implementation. Here, we write the smeared projector as $\tilde{P}_k = \sum_k p_{l_k} D_{l_k}$, where $D_{l_k}= h_{l_k} D(\zeta_k)$. For the input state $\rho_{\rm in}$, the projected state in the absence of ancilla noise reads $\tilde{P}_k\rho_{\rm in} \tilde{P}_k^\dag= \sum_{l_k l'_k} p_{l_k} p_{l'_k} D_{l_k} \rho_{\rm in} D_{l'_k}^\dag$. However, with the ancilla noise, the projection process changes into 
$\sum_{l_k l'_k} p_{l_k} p_{l'_k} e_{l_k l'_k} D_{l_k} \rho_{\rm in} D_{l'_k}^\dag$,
where $ e_{l_k l'_k} =\mathrm{exp}[-(\frac{\gamma_1}{2} + 2 \gamma_2) t_{l_k l'_{k} }]$ is the factor due to noise discussed in Eq. \eqref{Eq: uniformdecay} with $t_{l_k l_k'}$ being the evolution time for performing controlled operation for the implementation of $D_{l_k} \rho_{\rm in} D_{l'_k}^\dag$. Then, we discuss how to cancel the effect of $e_{l_k l'_k}$. Denoting $R=\sum_{l_k, l'_K } p_{l_k}  p_{l'_k} e_{l_k l'_k} ^{-1}$ and $p'_{l_k, l'_k} = p_{l_k} p_{l'_k} e_{l_k, l'_k}^{-1}/R$, we have
\begin{equation}
\tilde{P}_k\rho_{\rm in} \tilde{P}_k^\dag= R \sum_{l_k, l'_k} p'_{l_k, l'_k} e_{l_k, l'_k} D_{l_k} \rho_{\rm in} D_{l'_k}^\dag.
\end{equation}
Thus, by changing the sample probability for the indices $(l_k, l'_k)$ to $p'_{l_k, l'_k}$, we can again retrieve the projection by $\tilde{P}_k$. Note that the rescaling factor $R$ can be canceled with the normalization with the projection probability. Due to the rescaling factor $R$, the projection probability changes into:
\begin{equation}
q'= \mathrm{Tr}\bigg[\sum_{l_k, l'_k} p'_{l_k, l'_k} e_{l_k, l'_k} D_{l_k} \rho_{\rm in}  D_{l'_k}^\dag \bigg] =\frac{q}{R}
\end{equation}
for $q= \mathrm{Tr}\bigg[ \tilde{P}_k\rho_{\rm in} \tilde{P}_k^\dag \bigg]$. Therefore, the sampling overhead changes into $O(R^{2} q^{-2})$ from $O(q^{-2})$.

\end{document}